\documentclass[aip,amsmath,amssymb,reprint]{revtex4-2}
\usepackage{color}
\usepackage{hyperref}
\usepackage{enumitem}
\usepackage{graphicx,subfigure}
\usepackage[normalem]{ulem}
\usepackage{lineno}
\usepackage{booktabs}
\usepackage{balance}
\usepackage{array}
 \usepackage{multirow}
 \usepackage{hhline}
 \usepackage{makecell}
  \usepackage[symbol]{footmisc}
    \usepackage{siunitx} 
\newif\ifHighlitedChanges
\def\ifHighlitedChanges{\iftrue}
\ifHighlitedChanges
   
  \def\STRIKE#1{{\color{blue}\sout{#1}}}
\else
  
  \def\STRIKE#1{\relax}
\fi

\begin{document}
\title{Effects of quenched disorder on the kinetics and pathways of phase transition in a soft colloidal system}
\author{Gadha Ramesh}
\affiliation{Department of Physics, Indian Institute of Science Education and Research (IISER) Tirupati, Tirupati, Andhra Pradesh, 517619, India}
\author{Mantu Santra}
\affiliation{School of Chemical and Materials Sciences, Indian Institute of Technology Goa, Goa, 403401, India}
\author{Rakesh S. Singh}
\email{rssingh@iisertirupati.ac.in}
\affiliation{Department of Chemistry, Indian Institute of Science Education and Research (IISER) Tirupati, Tirupati, Andhra Pradesh 517507, India}

\begin{abstract}
Although impurities are unavoidable in real-world and experimental systems, most numerical studies on nucleation focus on pure (impurity-free) systems. As a result, the role of impurities in phase transitions remains poorly understood, especially for systems with complex free energy landscapes featuring one or more metastable intermediate phases. In this study, we employed Monte-Carlo simulations to investigate the effects of static impurities (quenched disorder) of varying length scales and surface morphologies on the nucleation mechanism and kinetics in the Gaussian Core Model (GCM) system, a model for soft colloidal systems. We first explored how the nucleation free energy barrier and critical cluster size are influenced by the fraction of pinned particles ($f_{\rm p}$) and the pinned cluster size ($n_{\rm p}$). Both the nucleation free energy barrier and critical cluster size increase sharply with increasing $f_{\rm p}$ but decrease as $n_{\rm p}$ grows, eventually approaching the homogeneous nucleation limit. On examining the impact of surface morphology on nucleation kinetics, we observed that the nucleation barrier significantly decreases with increasing the spherical pinned cluster (referred to as ``seed") size of face-centred cubic (FCC), body-centred cubic (BCC), and simple cubic (SC) structures, with BCC showing the greatest facilitation. Interestingly, seeds with random surface roughness had little effect on nucleation kinetics. Additionally, the polymorphic identity of particles in the final crystalline phase is influenced by both seed surface morphology and system size. This study further provides crucial insights into the intricate relationship between substrate-induced local structural fluctuations and the selection of the polymorphic identity in the final crystalline phase, which is essential for understanding and controlling crystallization processes in experiments.
\end{abstract}

\maketitle

\section{\label{sec:level1}Introduction}
Nucleation is a common phenomenon in nature, and the presence of impurities is inevitable in both real-world and experimental systems. Earlier theoretical studies have suggested that the presence of impurities can profoundly affect the nature of phase transitions~\cite{rounding_theory_1, rounding_theory_2, aizenberg_pnas_2018}. For instance, impurities in a system smooth the phase transition, often referred to as the ``rounding effect," causing the order parameter to vary continuously during the phase transition in a disordered system, whereas it is discontinuous in a pure system~\cite{rounding_theory_1, rounding_theory_2}. Recent computational studies also demonstrate a profound influence of the static impurities (or, quenched disorder) --- usually introduced via randomly pinning (or, freezing) a fraction of particles --- present in the system on the two-dimensional (2D) melting scenario~\cite{keim_2013, dijkstra_2015, ryzhov_2015, ryzhov_2021, ghoshal_pre}. However, despite the unavoidable presence of impurities in our daily life systems, majority of the theoretical and computational studies on nucleation focus on pure systems (for example, see Refs.~\cite{oxtoby1992homogeneous, pablo_book, frenkel_science, tanaka_pnas, pablo_ice_2015}), and consequently, the effects of impurities on nucleation and phase transitions are still not well understood.  

It is well-established that the presence of quenched disorder in a system can hinder its ability to crystallize~\cite{rpotherbhowmik2019effect, rpisingmandal2021nucleation, effectofpinning, rpotherbhowmik2016understanding, rpotherbrito2013jamming}. Therefore, many recent computational studies have utilized random pinning as a way to suppress crystallization and explore the relaxation behavior of deeply supercooled or glassy systems, aiming to test the validity of glass transition theories~\cite{rpglasschakrabarty2015dynamics, rpglasschakrabarty2015vanishing, rpglasschakrabarty2016understanding, rpglassdas2017pinning, rpglasskarmakar2013random, rpglasskob2013probing, rpglassliu2023pinning, rpglassozawa2015equilibrium, rpglassbhowmik2019particle, rpglassgokhale2014growing}. Majority of these studies, however, focus primarily on the sluggish relaxation dynamics of supercooled systems with random pinning and do not address the precise mechanisms by which quenched disorder suppresses the transition to a stable crystalline phase, and how it depends on the extent of quenched disorder present in the system. The role of length scales associated with the quenched disorder on the phase transition kinetics also remains largely unexplored. 

The complexity increases for materials with interaction potentials that feature multiple length and energy scales, which often results in a rich phase behavior characterized by several metastable phases, reflecting a complex free energy landscape~\cite{wales_book, mat_1, mat_2, mat_3, lekkerkerker_2002, chung_nat_phys, bagchi_2dm, bagchi_ss_2013, pablo_nature_2014, sciortino_natphys_2014, sleutel_nature_2018}. In these cases, it is important to consider not only the impact of impurities but also the role of intermediate phases (IPs) --- phases that exist between the metastable fluid and the globally stable solid phase. The presence of these IPs introduces the possibility of competition between thermodynamic factors (phase stability) and surface-induced kinetic factors (nucleation free energy) in controlling the phase transition. The situation becomes even more intriguing when the surface selectively promotes the formation of a particular IP polymorph. However, most earlier computational studies on heterogeneous nucleation have primarily focused on systems involving only two free energy basins (the metastable parent and the stable daughter phases), where kinetic facilitation is the main concern~\cite{cacciuto2004onset, sandomirski2011heterogeneous, lupi2014heterogeneous, lupi2016jcp, pedevilla2018heterogeneous, glatz2018heterogeneous, espinosa2019heterogeneous, lata2020multivalent, fitzner2020predicting, yuan2023rseeds, piaggi2024first}. As a consequence, the possibility of substrate-induced selection of IP polymorphs has received relatively less attention~\cite{davies2021routes, diaz2022template}.

The classical nucleation theory (CNT)~\cite{becker-doring-1935, frenkel-book-1955,pablo_book}, which effectively describes the mechanism and kinetics for two-phase systems (metastable and stable), often falls short in addressing phase transformations involving multiple intermediate phases. In such cases, one frequently observes non-classical scenarios like wetting-mediated or Ostwald step rule-like behavior~\cite{delhommelle_2011, yoreo_science_2014, sleutel_pnas_2014, bagchi_osr_2013, bagchi_ice_2014, kratzer_softmater_2015}. This suggests that the classical heterogeneous nucleation theory might also be insufficient in capturing the full range of phase transition scenarios in the presence of quenched disorder (static impurities or substrates; for example, see Refs.~\cite{fitzner2017pre, metya2021ice}). Numerous experimental and computational studies have already reported significant effects of impurity or seed surface morphology on nucleation kinetics and pathways, including the resulting crystal structure~\cite{fitzner2015many, smorphartusio2018surface, smorphbarron2017predicting, smorphhiranuma2014influence, davies2021routes, smorphwang2021morphology, grosfils2020impact, camarillo2024effect}.  For instance, in ice nucleation, the hydrophobicity or hydrophilicity of the surface can substantially alter the nucleation rate~\cite{cox2015molecular, glatz2018heterogeneous, cox2015molecular}. Since these studies provide insights into controlling crystal structure and kinetics through various substrate types, it is valuable to systematically investigate the impact of surface morphology and seed size on heterogeneous nucleation mechanisms and kinetics. 
 
In this work, we used the Gaussian Core Model (GCM)~\cite{gcm} to computationally explore the effects of quenched disorder, with varying length scales and surface morphologies, on nucleation mechanism and kinetics. This model exhibits a rich phase behavior with fluid-FCC, fluid-BCC and FCC-BCC phase coexistence lines~\cite{meijer_1997}, making it particularly suitable for studying the influence of substrates on nucleation kinetics and polymorph selection. We systematically varied the size and surface morphology of the quenched disorder (\textit{i.e.}, pinned clusters/seeds) to examine their impact on the nucleation free energy barrier and phase transformation pathways. Our results reveal that nucleation kinetics and polymorph selection are significantly influenced by the size and surface morphology of the impurities. Additionally, we investigated how the equilibrium structural order of interfacial fluid particles can provide insights into polymorph selection.

The rest of this paper is organized as follows. Section~\ref{sec:method} details the computational protocol followed for simulations of the soft colloidal system modelled via the GCM potential~\cite{gcm}. In Section~\ref{subsec:1} we discuss the effects of random quenched disorder of varying length scales on the nucleation free energy barrier. The surface-induced crossover from the homogeneous to heterogeneous nucleation for different seed sizes and surface morphologies is discussed in Section~\ref{subsec:2}. The surface-induced polymorph selection and a detailed analysis of the microscopic mechanism of polymorph selection and finite-size effects are discussed in Sections~\ref{subsec:3} and~\ref{subsec:4}, respectively. The signatures of the surface-induced polymorph selection hidden in the equilibrium structural fluctuation of the metastable fluid is discussed in Section~\ref{subsec:6}. Section~\ref{conclusions} summarizes the major conclusions from this work. 

\section{\label{sec:method}Simulation methods}
\subsection{Model details}
We performed Monte-Carlo (MC) simulations~\cite{frenkel_book} on a system interacting via the GCM potential~\cite{gcm}, $u(r) = \epsilon \exp\left[-(r/\sigma)^2\right]$, where $\sigma$ is the particle diameter and $\epsilon$ is the energy when the inter-particle separation ($r$) is zero. $\sigma$ and $\epsilon$ are used as units of length and energy, respectively. In this study, we have truncated the interaction potential at a distance $r_c = 4.0\sigma$ and shifted to zero. The pressure ($P$)-temperature ($T$) phase diagram of the GCM model system is well-reported~\cite{tanaka_soft_2012, santi_gcm_2005} and is shown in Fig.~\ref{fig1}A).

\subsection{Computation of chemical potential}\label{subsec:phase}
\begin{figure}
     \includegraphics[scale=0.4]{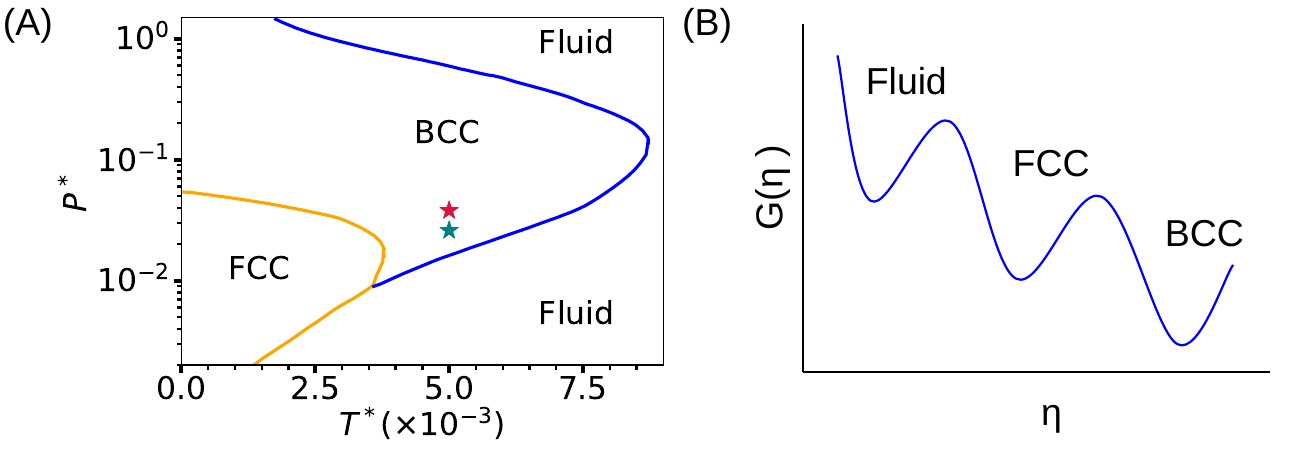}
    \caption{(A) The phase diagram of the Gaussian Core Model (GCM) system reproduced from Refs.~\cite{tanaka_soft_2012, santi_gcm_2005} is shown. The green and red asterisks, referred to as SP1 and SP2, respectively, represent the thermodynamic state points studied in this work. (B) A schematic representation of the free energy surface at these thermodynamic conditions is shown. Note that, BCC is the globally stable phase, and FCC is metastable with respect to the BCC and stable with respect to the fluid phase at both the state points. In this work, the FCC is referred to as intermediate phase (IP).}
    \label{fig1}
\end{figure}

In the phase diagram we chose two state points, named as SP1 ($T^{\ast}=0.005, P^{\ast}=0.026$) and SP2 ($T^{\ast}=0.005, P^{\ast}=0.038$), to perform our study (see Fig.~\ref{fig1}A). Here, $\ast$ indicates reduced units. At these state points BCC is the globally stable phase and the fluid and FCC phases exist as a metastable state. To estimate the nature of the underlying global free energy and the chemical potential difference between the fluid and the solid (BCC and FCC) phases (denoted as, $\Delta\mu_{\rm BCC}$ and $\Delta\mu_{\rm FCC}$, respectively), we carried out free energy calculations employing umbrella sampling technique using the order parameter introduced in Ref.~\cite{comp_mu_2012}. The chemical potential of the BCC and the FCC phases with respect to the liquid phase is reported in Table~\ref{tab:table1}.
\subsection{Identification of solid-like particles and polymorphs}\label{subsec:polymorph}
Solid-like crystallites in the metastable fluid phase are identified using the method introduced by Frenkel and co-workers~\cite{frenkel_1996, frenkel_2005}. This method first identifies the local bond-orientational symmetry of particle $i$ using a complex vector $q_{lm}(i)$~\cite{bop} as ,
\begin{equation}\label{eq:1}
 q_{lm}(i) = \frac{1}{N_b(i)}\sum_{j=1}^{N_b(i)}Y_{lm}(\mathbf{r_{ij}})
\end{equation}
where $N_b(i)$ is the number of nearest neighbors of the $i^{th}$ particle. Two particles are considered to be neighbors if the distance between them ($|\mathbf{r_{ij}}|$) is less than the radial cut-off distance of $q_c=1.38/(\rho \sigma^3)^{1/3}$, where $q_c$ is the (average) radius of the first shell of fluid particles (measured as the position of the minimum separating the first and second peaks in the radial distribution function), $\rho \sigma^3$ is the reduced density. $Y_{lm}(\mathbf{r_{ij}})$ is the spherical harmonics and $\mathbf{r_{ij}}$ is the distance vector between the particle $i$ and its neighbor $j$. $l$ and $m$ are integers with $-l \le m \le l$. The unit vector of $q_{lm}(i)$ is given by,
\begin{equation}
d_{lm}(i) = \frac{q_{lm}(i)}{\Big(\sum_{m=-l}^{m=l} |q_{lm}(i)| ^2\Big)^{1/2}}.
\end{equation}
Using the unit vector $d_{lm}(i)$, a scalar product $S_l(i,j)$ which measures the correlation in bond orientational order between neighboring particles can be defined as,
\begin{equation}
S_l(i,j) = \sum_{m=-l}^{m=l} d_{lm}(i).d_{lm}^*(j),  
\end{equation}
where the superscript $^*$ indicates complex conjugate. Two neighboring particles $i$ and $j$ are considered to be connected if $S_6(i,j) > 0.7$. The particle $i$ is identified as solid-like if the number of such connections is more than $7$. A solid-like cluster is defined as the collection of solid-like particles sharing a common neighborhood defined by the radial cut-off distance of $q_c$. 

To assign the polymorphic identity of a solid-like particle, we employed the locally averaged bond orientational order parameter introduced by Lechner and Dellago~\cite{lechner_jcp_2008}. Using $q_{lm}(i)$ given in Eq.~\ref{eq:1}, one can define a locally averaged complex vector $\bar{q}_{lm}(i)$ as,
\begin{equation}\label{eq:2}
 \bar{q}_{lm}(i) = \frac{1}{N_b(i) + 1}\sum_{j=0}^{N_b(i)} q_{lm}(j)
\end{equation}
where $j=0$ indicates the particle $i$ itself. Given the coarse-grained complex vector $\bar{q}_{lm}(i)$, one can further define coarse-grained order parameters  $\bar{q}_{l}(i)$ and $\bar{w}_{l}(i)$ as
\begin{equation}\label{eq:3}
 \bar{q}_l(i) = \sqrt{\frac{4\pi}{2l+1}\sum_{m=-l}^{l}|\bar{q}_{lm}(i)|^2}
\end{equation}
and 
\begin{equation}\label{eq:4}
\bar{w}_l(i) = \sum_{m_1 + m_2 + m_3 = 0} \left(\begin{array}{clcr}
l & l & l \\
m_1    & m_2 & m_3                                        
\end{array}\right) \frac{\bar{q}_{lm_1}(i)\bar{q}_{lm_2}(i)\bar{q}_{lm_3}(i)}{\left[\sum_{m=-l}^{l}|\bar{q}_{lm}(i)|^2\right]^{3/2}}
\end{equation}
where the term in the parentheses $\left( ... \right)$ indicates the Wigner $3j$ symbol. The integers $m_1$, $m_2$ and $m_3$ range from $-l$ to $l$ and only the terms with $m_1+m_2+m_3=0$ are allowed to contribute to the summation. Once $\bar{w}_l(i)$ is defined, we identify a previously assigned solid-like particle as BCC-like if $\bar{w}_6 > 0$, whereas it is considered to be FCC-like if $\bar{w}_6 \le 0$~\cite{tanaka_2012}. Note, in this work, we have not made a distinction between the FCC- and HCP-like particles.   
\begin{table}
\caption{The chemical potential difference (in reduced units) between the metastable fluid and stable (BCC and FCC) solid phases (denoted by $\Delta \mu_{\rm fcc}$ and $\Delta \mu_{\rm bcc}$, respectively) at SP1 and SP2 is reported.}
\label{tab:table1}
\begin{tabular}{c|l|l|l|l}
\hline
State Point & $T^{\ast}$ & $P^{\ast}$ & $\Delta \mu_{\rm fcc} $ & $\Delta \mu_{\rm bcc} $ \\
\hline
SP1 &  $0.005$  & $0.026$   & $-0.144$         & $-0.180$         \\
SP2 &  $0.005$  & $0.038$   & $-0.274$         & $-0.321$        
\end{tabular}
\end{table}
 \subsection{Nucleation free energy barrier computation}
 The fluid to solid nucleation free energy profiles are computed by employing umbrella sampling method~\cite{umbrella} with size of the largest solid-like cluster as the order parameter in $NPT$ ensemble consisting of $N = 6000$ particles. The force constant of the umbrella potential was taken to be $\lambda=0.1 k_{\rm B}T$.
\begin{figure*}[ht!]
     \includegraphics[width=0.8\linewidth]{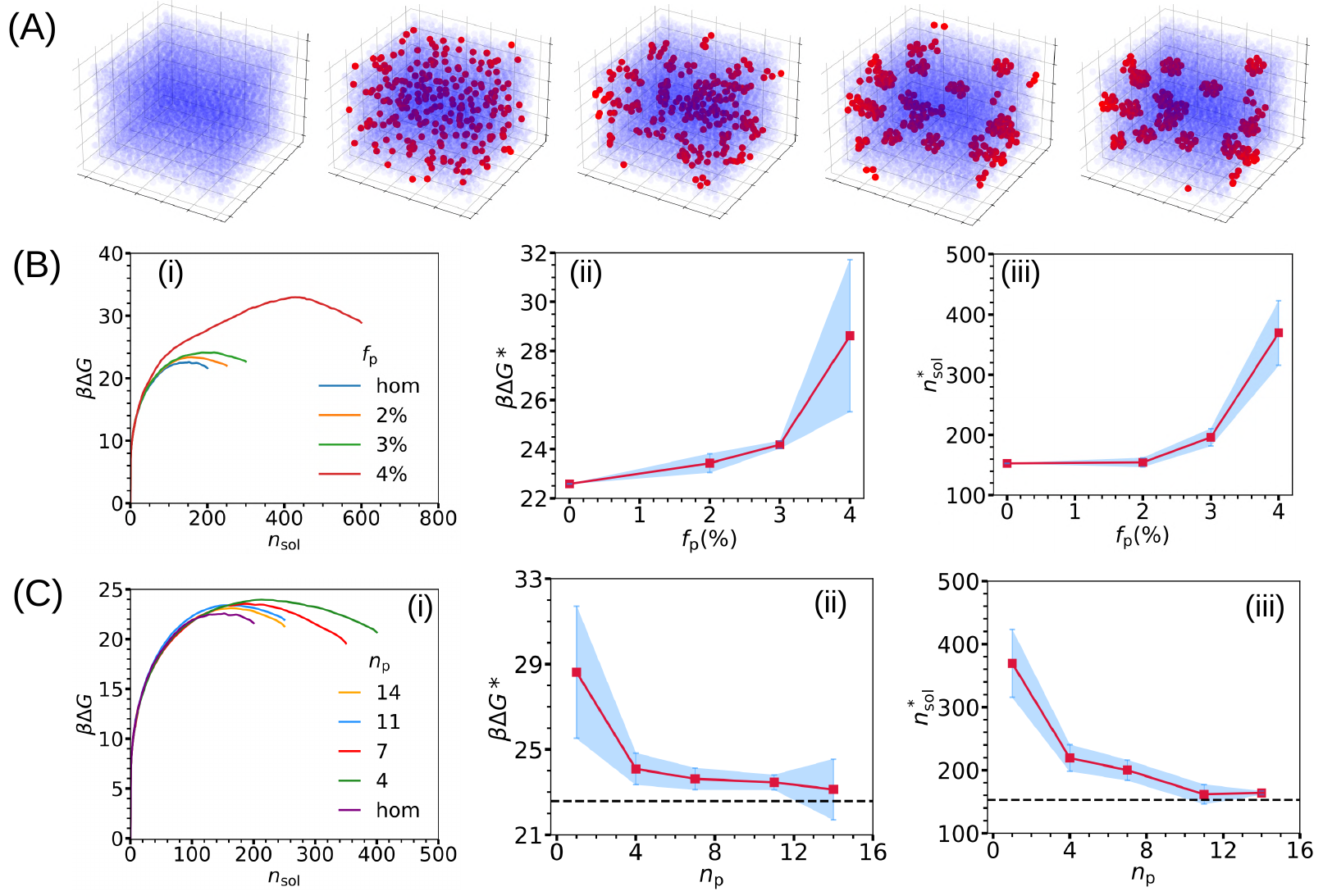}
    \caption{(A) Snapshots of the homogeneous and inhomogeneous systems with randomly pinned particles/clusters of different sizes ranging from single-particle to the roughly first coordination shell are shown. The system contains $N = 6000$ particles and $4\%$ of the total particles are pinned in the system. (B) The dependence of (i) the nucleation free energy surface ($\beta \Delta G$), (ii) nucleation free energy barrier ($\beta \Delta G^*$), and (iii) the critical cluster size ($n_{\rm sol}^*$) for the  fluid to solid transition on the fraction of pinned particles in the system ($f_{\rm p}$) at SP2 is shown. Here, ``hom" indicates the homogeneous case. (C) The dependence of $\beta \Delta G$ (i), $\beta \Delta G^*$ (ii), and $n_{\rm sol}^*$ (iii) on the size of the pinned regions ranging from $n_{\rm p} = 1$ (single-particle or point pinning) to $n_{\rm p} = 14$ (roughly upto the first coordination shell) for $f_{\rm p} = 4\%$ is shown. The dashed black line represents the same for the homogeneous system case. The shaded blue region is the error bar concerning three independent simulations with different realizations of the pinned particle spatial distribution.} 
    \label{figure2}
\end{figure*} 
\begin{figure*}[t]
    \centering
    \includegraphics[width=0.84\linewidth]{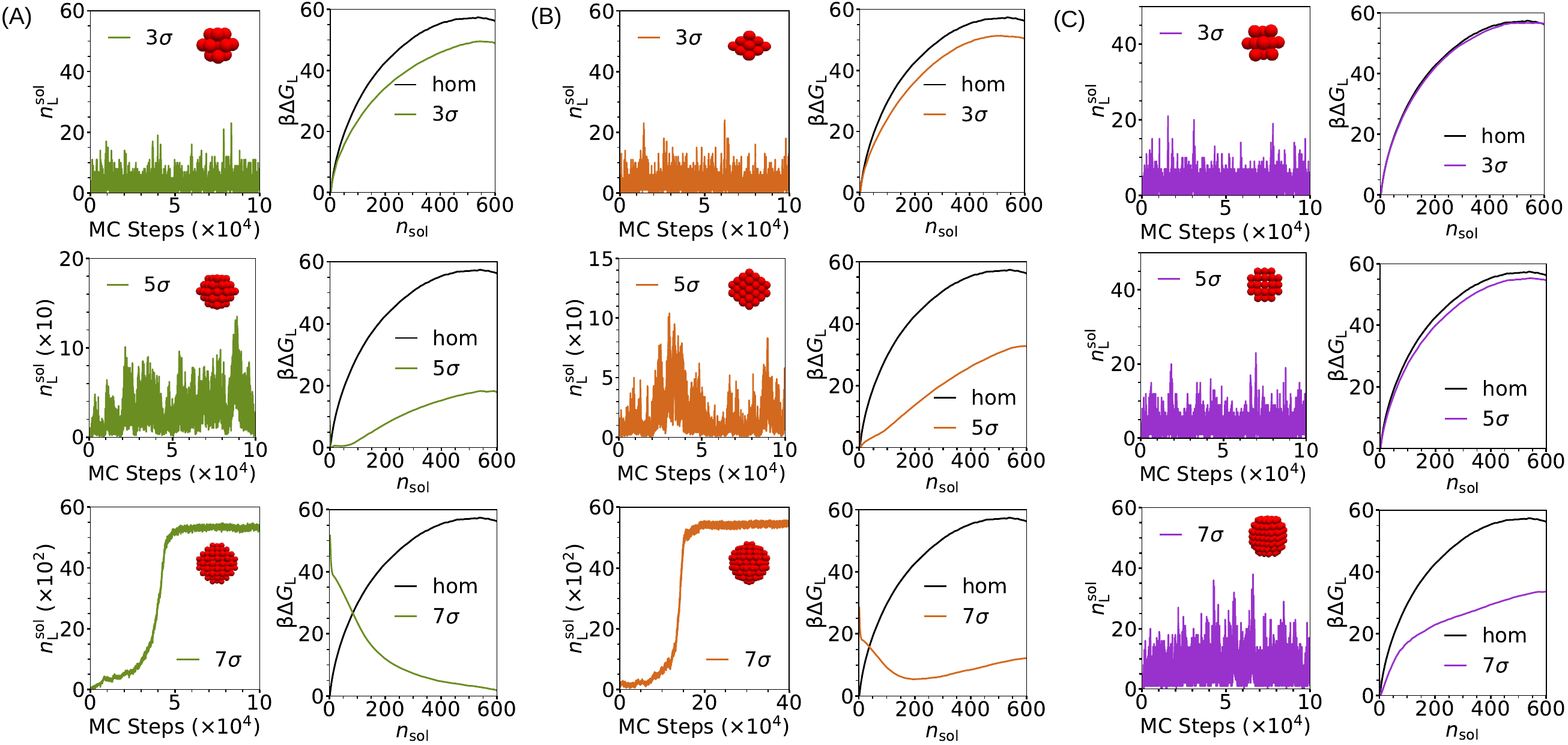}
    \caption{(A) The variation of the largest solid-like cluster ($n_{\rm L}^{\rm sol}$) with the MC steps (left) along with the nucleation free energy profile for the largest solid-like cluster ($\beta \Delta G_{\rm L}$) is shown for the system containing $N = 6000$ fluid particles with a spherical BCC (A), FCC (B) and SC (C) seed at SP1. Here, ``hom” indicates the nucleation free energy profile for the homogeneous case. We have shown the results for three different seeds of radii of $3 \sigma$, $5 \sigma$, and $7 \sigma$. We note a crossover from the activated to the barrierless phase transition on increasing the size of the FCC and BCC pinned seed. However, for the SC seed, we observe a significant lowering of the nucleation barrier $\beta \Delta G_{\rm L}^*$ but not a downhill (or, barrierless) transition on increasing the seed size.} 
    \label{figure3}
\end{figure*}
\begin{figure}[t!]
    \includegraphics[scale=0.125]{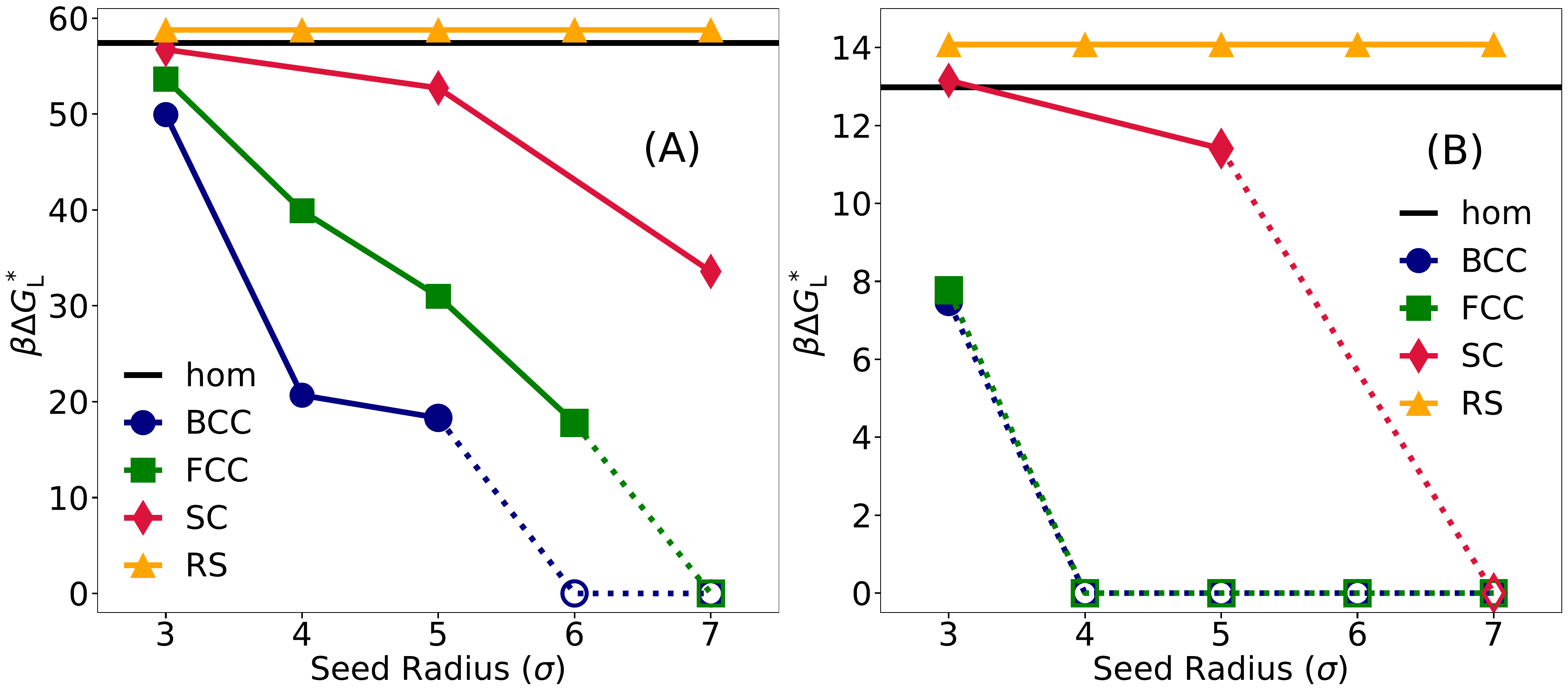}
    \caption{The free energy barrier of nucleation for the largest solid-like cluster, $\beta \Delta G_{\rm L}^*$, for different seed types and sizes at SP1 (A) and SP2 (B) is shown. The solid black line indicates the $\beta \Delta G_{\rm L}^*$ for the homogeneous nucleation case (denoted as ``hom"). The dotted lines with open markers demonstrate the seed sizes for which the phase transition is downhill or barrierless. The system contains $N = 6000$ particles.}
    \label{figure4}
\end{figure}
\section{\label{sec:level2}Results and discussions}
\subsection{\label{subsec:1} Nucleation kinetics in the presence of random quenched disorder of different length scales}
To probe the sensitivity of crystal nucleation kinetics from the metastable fluid on the presence of random quenched disorder of varying length scales, we computed the nucleation free energy profile of solid-like particles by varying both the fraction and size of pinned particles/clusters. We pinned clusters of $1$ (referred to as single-particle or point pinning), $4$, $7$, $11$, and $14$ particles (approximately up to the first coordination shell). To create a pinned cluster of $n_{\rm p}$ particles, we first randomly selected a particle and its $n_{\rm p} - 1$ nearest neighbors, and then froze their movement throughout the simulation. Figure~\ref{figure2}A shows representative snapshots of sample initial configurations for both homogeneous (unpinned) and pinned systems with a fixed fraction of pinned particles. It is important to note that the pinned and unpinned particles are identical, except that the pinned particles remain immobile throughout the simulation.  

We first investigated the sensitivity of the nucleation free energy barrier for solid-like clusters to the fraction of point pinned particles ($f_{\rm p}$). Since the pinned particles are randomly distributed in space, we performed at least three independent simulations, each with a different realization of the initial pinned configuration. In Fig.~\ref{figure2}B(i), we show the nucleation free energy profile [$\beta \Delta G$ vs. the size of the solid-like cluster ($n_{\rm sol}$)] for various $f_{\rm p}$ ranging from $0\%$ (the homogeneous or unpinned limit) to $4\%$. As is evident from the figure, the nucleation free energy profile is highly sensitive to $f_{\rm p}$, with both the nucleation barrier ($\beta \Delta G^*$) and the critical cluster size ($n_{\rm sol}^*$) increasing sharply as $f_{\rm p}$ increases (see Figs.~\ref{figure2}B(ii) and~\ref{figure2}B(iii)). We have not presented the nucleation profile for $f_{\rm p} \geq 5\%$ as we could not achieve a converged profile due to the sluggish structural relaxation of the metastable fluid containing solid-like clusters.  

In a recent study, Quigley and coworkers~\cite{rpisingmandal2021nucleation} examined how impurities affect droplet free energy in a 2D Ising model by introducing non-magnetic impurities that do not interact with spin states. They found that both the free energy barrier and critical cluster size decrease as impurity fraction increases. Interestingly, our results show the opposite: both the nucleation barrier and critical cluster size increase with $f_{\rm p}$. This indicates that the impact of quenched disorder on phase transition kinetics depends on the nature of impurity interactions with the surrounding (metastable) phase particles and the degree of frustration they introduce in the growing stable phase droplets. 
  
We further investigated the dependence of the nucleation free energy barrier on the length scale of inhomogeneities (quenched disorder) by randomly freezing regions of varying sizes ($n_{\rm p}$) in the system,  while keeping the fraction of pinned particles constant at $4\%$. The only change was the spatial localization of the pinned particles as $n_{\rm p}$ varied (see Fig.~\ref{figure2}A). In Fig.~\ref{figure2}C(i), we show the dependence of the nucleation profile on the size of the pinned cluster for a fixed fraction of pinned particles. As is evident from the figure, the nucleation profile is delicately sensitive to the length scale associated with quenched disorder in the system. As the size of the pinned cluster increases, both the nucleation free energy barrier $\beta \Delta G^*$ and the critical cluster size $n_{\rm sol}^*$ decrease (Figs.~\ref{figure2}C(ii) and~\ref{figure2}C(iii)) and gradually approach their corresponding homogenous limits. 
\begin{figure*}[t]
    \centering
    \includegraphics[width=0.84\linewidth]{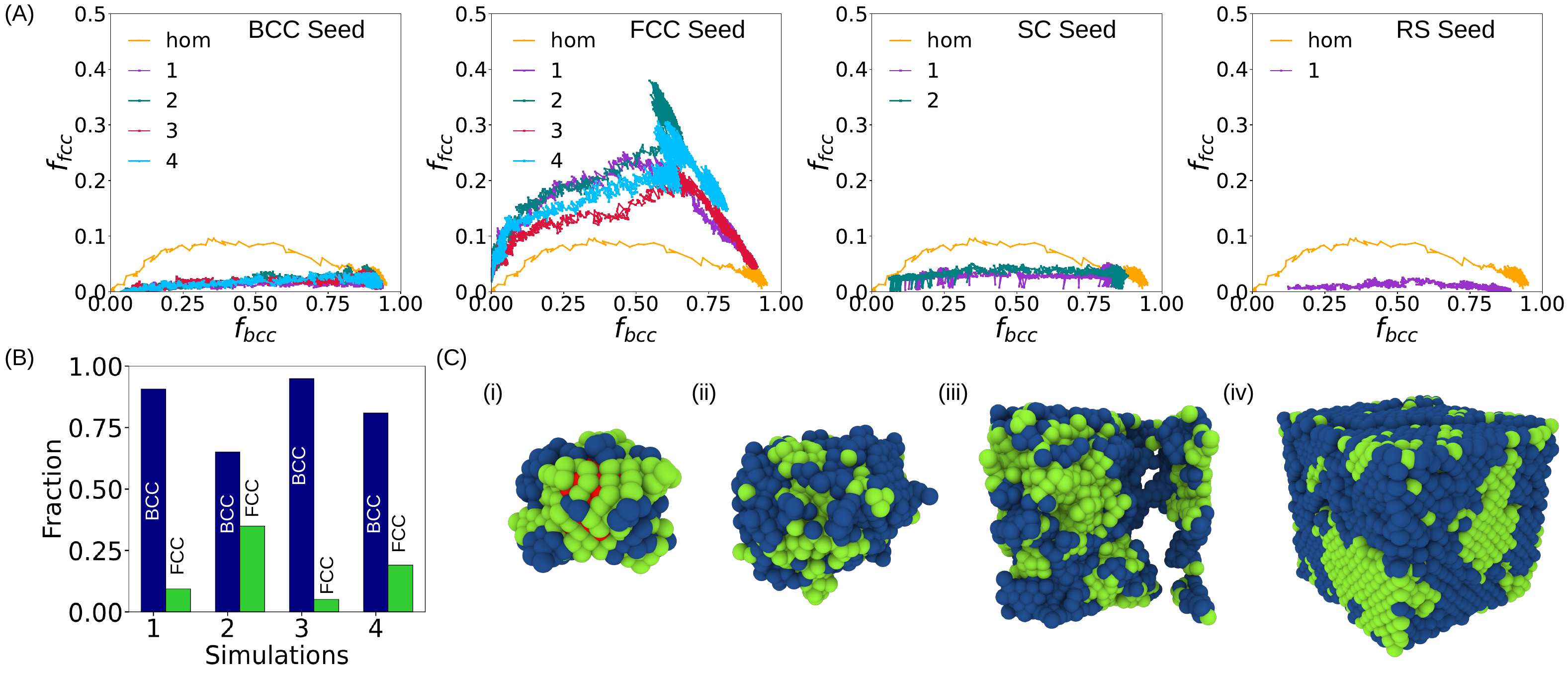}
    \caption{(A) The phase transition pathways of the metastable fluid phase in the two-dimensional --- fraction of the BCC ($f_{\rm bcc}$) and FCC ($f_{\rm fcc}$)-like particles --- space at SP1 for different types of seeds of radius $7 \sigma$ is shown. The system  consists of $N = 10000$ particles. The orange line indicates the phase transition path for the homogeneous system case. (B) The fraction of the FCC- and BCC-like polymorphs in the final solid phase formed after the phase transition for four independent configurations of the metastable fluid containing an FCC seed of radius $7 \sigma$. (C) Representative snapshots depicting the polymorphic identity of all the particles at different stages of the growth trajectory (Simulation $2$ in Fig.~\ref{figure5}B) are shown in (i)-(iv). The blue spheres represent the BCC-like, and green spheres represent the FCC-like particles in the system. The seed is represented in red color.}
    \label{figure5}
\end{figure*}  
\subsection{\label{subsec:2} Surface-induced facilitation of nucleation kinetics}
Despite observing a decrease in the nucleation barrier with increasing pinned cluster size in the previous section, we did not detect a crossover from homogeneous to heterogeneous nucleation, where nucleation kinetics would be facilitated compared to the homogeneous case. To gain deeper insights into the impact of surface morphology on nucleation kinetics, we investigated how the surface characteristics of the pinned cluster influence the process. Specifically, we pinned a single spherical cluster of particles (referred to as a ``seed") with varying surface morphologies --- FCC, BCC, simple cubic (SC), and an amorphous-ordered rough sphere (RS) --- at the center of the simulation box. For the RS seed, we first selected a central particle in the thermally equilibrated fluid, and all particles in its neighborhood within a chosen radius, including the tagged particle, were pinned. This created a rough surface morphology, allowing us to study the effects of surface roughness on phase transition kinetics. To explore the different realizations of surface roughness, we performed the nucleation barrier computations using at least three independent simulations, each with a different RS seed. For all cases, the radius of the spherical seed was varied between $3\sigma$ and $7\sigma$. 

To investigate the surface-induced effects on nucleation kinetics, we first computed the size fluctuations and growth of the largest solid-like cluster ($n_{\rm L}^{\rm sol}$) along with the nucleation free energy profile for the largest solid-like cluster ($\beta \Delta G_{\rm L}$) for different seed types (see Fig.~\ref{figure3}) at SP1. The $\beta \Delta G_{\rm L}$ provides an estimate of the actual (kinetic) stability limit of the metastable system and depends on system size~\cite{mantu_prl, jstat_singh}. It is evident from the figure that the nucleation barrier decreases significantly as the seed size increases for the FCC, BCC, and SC seeds. Notably, nucleation kinetics is facilitated more by the BCC seed compared to the FCC and SC seeds. For the BCC and FCC seed cases, we observe a seed-size dependent crossover from activated to a barrierless (heterogeneous) nucleation, where the phase transition occurs completely downhill. In contrast, the SC seed does not reduce the barrier as much as the BCC and FCC seeds, likely because the SC seed does not contain either the stable BCC-like or the intermediate FCC-like surface morphologies. We further observe that the RS seed doesn't facilitate nucleation kinetics, even with a seed radius of $7\sigma$ (see Fig.~\ref{figs1} in the Supplementary Materials).  

We also explored the impact of the underlying free energy landscape, particularly the stability of different fluid and solid phases, on surface-induced nucleation kinetics facilitation. For this, we computed the seed-size-dependent nucleation free energy barrier for the largest cluster ($\beta \Delta G_{\rm L}^*$) for various seed types at SP1 and SP2 (see Fig.~\ref{figure4}). At SP2, the fluid is more metastable (\textit{i.e.}, less stable) with respect to the BCC and FCC phases compared to SP1 (see Table~\ref{tab:table1}). We observed that the onset of the crossover from activated to barrierless phase transition --- or, the critical seed size that triggers a barrierless transition --- is highly sensitive to thermodynamic conditions. For a given seed type, when the fluid phase is more metastable (at SP2), even a smaller seed leads to a barrierless phase transition. Interestingly, the RS seed consistently results in a slightly higher nucleation barrier than in the homogenous case, regardless of seed size. These findings indicate that both the size and surface morphology of static inhomogeneities significantly influence nucleation kinetics facilitation. 
\begin{figure*}
    \includegraphics[scale=0.6]{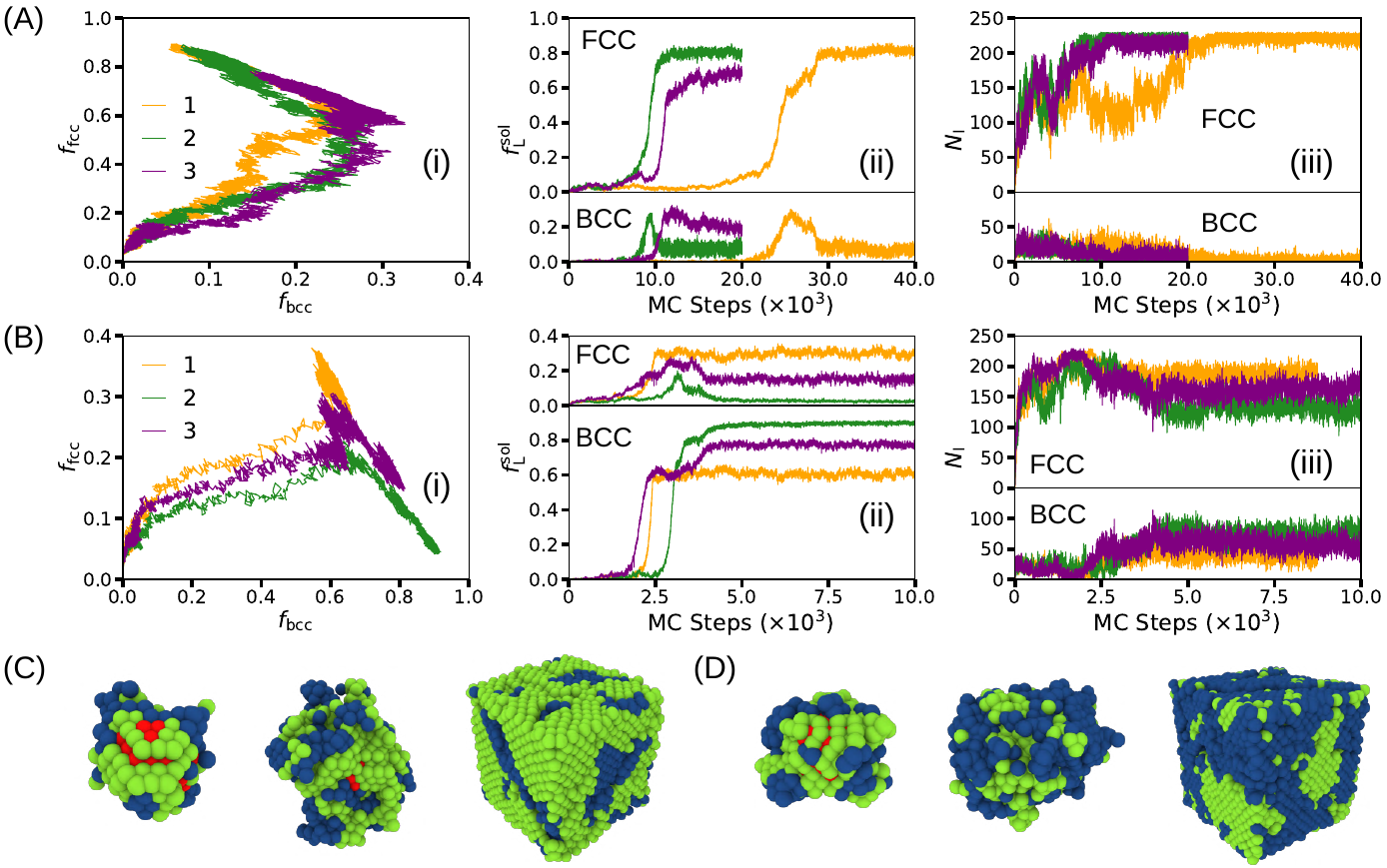}
    \caption{(A) The phase transition pathways of the metastable fluid phase containing an FCC seed of radius $7 \sigma$ for $N=6000$ in the $f_{\rm bcc} - f_{\rm fcc}$ plane at SP1 for two independent initial configurations is shown (i). The evolution of (ii) the fraction of the FCC- and BCC-like particles in the largest solid-like cluster ($f_{\rm L}^{\rm sol}$), and (iii) the number of FCC- and BCC-like interfacial particles ($N_{\rm I}$) with the MC steps is also shown. (B) The same is shown for $N=10000$. The snapshots of system at different stages during the transition to the FCC-dominated final configuration for $N=6000$ (C), and to the BCC-dominated final configuration for $N=10000$ (D) are shown. The FCC seed is represented in red color, the BCC-like particles in blue, and the FCC-like particles in green. During the initial growth, the nucleating particles are predominantly FCC-like for both the system sizes.}
    \label{figure6}
\end{figure*}
\subsection{\label{subsec:3} Surface-induced polymorph selection}
Here, we have elaborately probed how the morphology of the seed surface influences the selection of the polymorphic identity of particles in the final crystalline phase. We note that, at the thermodynamic conditions used in this work (SP1 and SP2), the BCC phase is globally stable, while the FCC phase is metastable relative to the BCC phase with varying degrees of metastability (see Table~\ref{tab:table1}). Consequently, the fluid phase could potentially crystallize into either of these two (BCC and FCC) phases. Previous studies have shown that the morphology and roughness of the seed surface can impact the nucleation pathway, thereby influencing the selection of the final crystalline polymorph~\cite{spolycurcio2008probabilistic, davies2021routes, spolydiaz2022template, spolyyao2022anisotropic, spolyzhang2023controlling}. Therefore, for a given seed size, it is important to examine how the seed type (specifically, its surface morphology) can control the polymorphic identity of the particles in the resulting crystalline structure. 

Figure~\ref{figure5}(A) illustrates the phase transition pathways of the metastable fluid phase influenced by different seed types of size $7\sigma$ for $N=10, 000$. With the seed size fixed at $7\sigma$, we have projected the transition path onto a two-dimensional plane --- the fraction of BCC-like ($f_{\rm bcc}$) and FCC-like ($f_{\rm fcc}$) particles in the largest solid-like cluster --- for all seed types at SP1. We observe that the BCC, SC and RS seed types lead to a transition to the globally stable BCC phase (see Fig.\ref{figure5}A). In all these cases, rapid growth of the BCC-like cluster is observed, while $f_{\rm fcc}$ remains very low throughout the transition phase across all independent trajectories. This indicates that the fluid particles predominantly transition to the BCC phase, with only a small fraction converting to the FCC phase.

However, for the FCC seed, competition between the stable BCC phase and the intermediate FCC phase emerges (see Figs.~\ref{figure5}A). In Fig.~\ref{figure5}B we report the polymorphic identity of different particles (FCC- or BCC-like) in the final crystalline configuration for four independent transition trajectories of the metastable fluid with FCC seed of radius $7\sigma$ at SP1. It is evident that, for the FCC seed, some of the final configurations contain a considerable fraction of FCC-like particles, with the dominant phase being BCC. In all cases, nucleation is initiated at the surface of the FCC seed, with the particles near the surface predominantly exhibiting FCC-like local structural features (see Fig.~\ref{figure5}C). During the phase transition, the fluid particles transform into BCC-like and FCC-like solids, leading to competitive growth between these two phases in few trajectories. After this competitive growth phase, the system crystallizes into a structure with a dominant BCC polymorph with a significant fraction of FCC-like particles. 

We also note that the signatures of heterogeneous nucleation are observed for the SC seed case (as reported in the previous section, Fig.~\ref{figure3}C). Solid nucleation begins on the surface of the seed with BCC-like particles crystallizing onto it. However, unlike the BCC seed, where crystallization occurs over the entire seed surface, in this case, only a small portion of the seed surface promotes growth (see Fig.~\ref{figs2} in the Supplementary Materials). Additionally, no competing nucleation and growth of BCC and FCC phases are observed. Unlike the crystalline (FCC, BCC, and SC) seeds, the amorphous RS seed does not give rise to any facilitation and the polymorphic identity of the particles in the final solid is predominantly BCC. We also investigated the surface-induced  polymorphic identity selection for different types of seeds at SP2 and found a similar trend (see Fig.~\ref{figs3} in the Supplementary Materials). Thus, the seed surface morphology and size play a crucial role in facilitating nucleation and even determining the polymorphic identity of particles in the final crystalline phase. 
\subsection{\label{subsec:4} Finite-size effects on the surface-induced polymorph selection}
It has already been reported that finite system size in computational studies significantly impacts crystal nucleation, primarily due to unphysical interactions between crystalline nuclei and their periodic images~\cite{hussain2021quantify, hussain2022quantify}. However, the effects of finite size on polymorph selection during crystallization, particularly in the case of heterogeneous nucleation, remain poorly understood. To investigate these finite-size effects on surface-induced phase transition and polymorph selection, we compared the transition pathways in the $f_{\rm bcc} - f_{\rm fcc}$ plane for two system sizes, $N=6000$ and $N=10000$ with an FCC seed of radius $7\sigma$ at SP1 (Figs.~\ref{figure6}A and~\ref{figure6}B). For $N=10000$, all three independent simulations nucleate into the stable BCC phase with a small FCC fraction, even though competitive growth between the FCC and BCC phases is observed during the initial growth phase (discussed in the previous section and in Fig.~\ref{figure6}B(i)). In contrast, for $N=6000$, the FCC seed plays a more prominent role in facilitating FCC (which is metastable with respect to BCC) polymorph selection. All three independent trajectories result in the FCC phase, with a small fraction of BCC-like particles, following a competitive growth between these two phases during the early stages (see Fig.~\ref{figure6}A(i)). This indicates that finite-size effects do indeed influence the phase transition pathway, and consequently, the selection of polymorphs.
\begin{figure*}[ht!]
    \centering
    \includegraphics[scale=0.37]{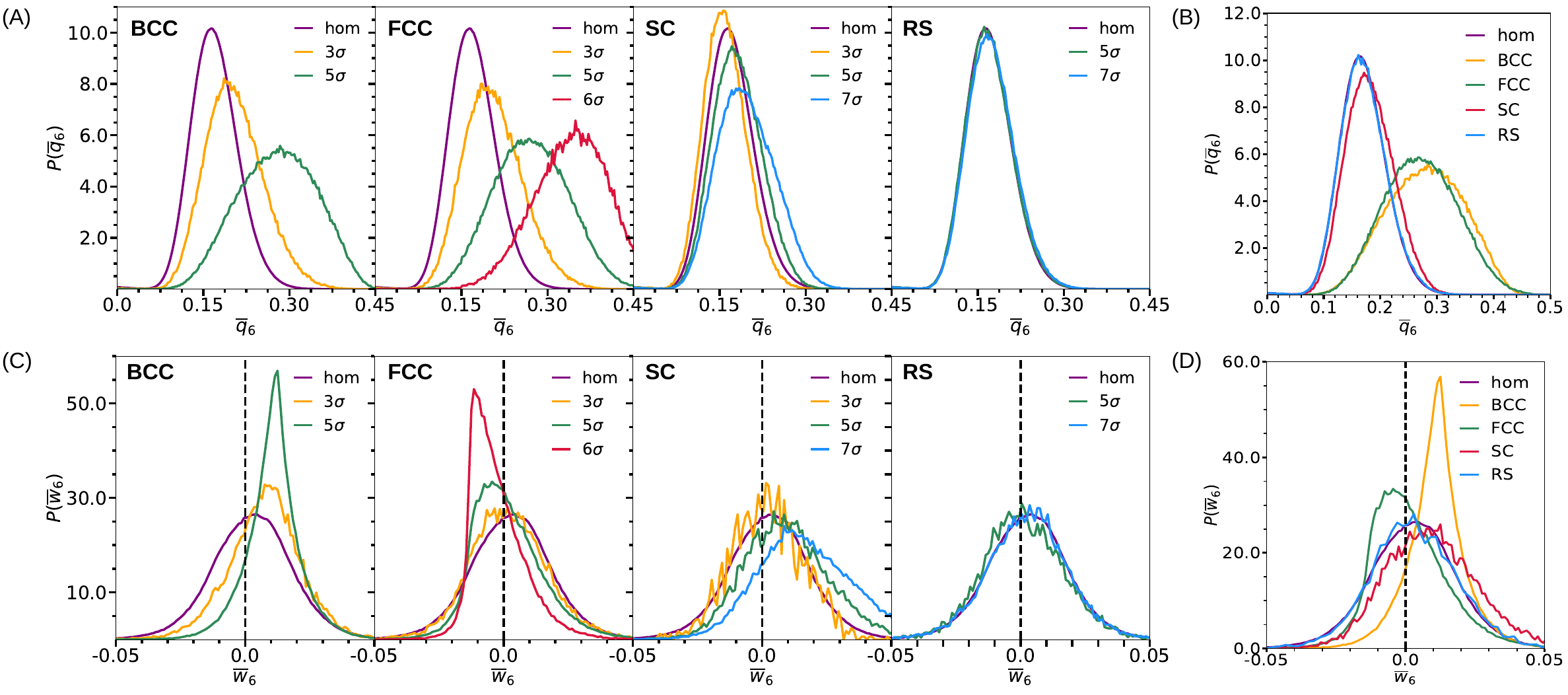}
    \caption{(A) The dependence of the local coarse-grained bond orientational order parameter $\bar{q_6}$ (Eq.~\ref{eq:3}) distribution of the fluid particles in the vicinity (within a distance of $2 \sigma$ from the seed surface) of different types and sizes of the seed at SP1. (B) The local coarse-grained bond orientational order parameter $\bar{q_6}$ distributions of fluid particles in the vicinity of different  seeds of radius $5 \sigma$. (C) The dependence of the $\bar{w_6}$ (Eq.~\ref{eq:4}) distribution of partly crystalline fluid particles (fluid particles with $\bar{q_6} > 0.25$) in the vicinity of the different seeds. The same is shown for $\bar{w_6}$. (D) The $\bar{w_6}$ distributions of partly crystalline fluid particles around different seeds of radius $5 \sigma$. The system contains $N = 6000$ particles.}
    \label{figure7}
\end{figure*} 

We also computed the evolution of the fraction of BCC- and FCC-like particles in the largest solid-like cluster ($f_{\rm L}^{\rm sol}$) along with the evolution of the polymorphic identity of the interfacial fluid particles $N_{\rm I}$ (see Figs.~\ref{figure6}(ii) and (iii)). Interfacial particles are identified as those having at least one seed particle as their nearest neighbor. During the transition, both BCC- and FCC-like particles grow competitively, with a slightly higher fraction of FCC-like particles for the $N = 6000$ system and a slightly higher fraction of BCC-like particles for the $N = 10000$ system (see Figs.~\ref{figure6}A(i) and~\ref{figure6}B(i)). After this initial growth phase, FCC growth dominates over the BCC growth in the $N = 6000$ system, while the opposite trend is observed for $N = 10000$. Interestingly, in both systems sizes, the interfacial  particles are predominantly FCC-like with a slightly higher fraction for $N = 6000$ compared to $N = 10000$ (see Fig.~\ref{figure6}(iii) and the representative snapshots in Figs.~\ref{figure6}C and~\ref{figure6}D). Additionally, for the $N = 10000$ system, we note that the BCC phase growth also originates at the surface of the seed, rather than in the bulk fluid (see Fig.~\ref{figs4} in the Supplementary Materials).  

This finite-size effect can be understood in terms of an interplay between surface-facilitated FCC (or, IP)-like ordering and the (globally) stable BCC-like ordering within the metastable fluid. For a given system size, the extent of surface-induced local structural changes depends on the seed size. Therefore, in finite-sized systems (\textit{e.g.}, confined systems), the degree of surface-induced local structural ordering --- and consequently, the growth of the surface-favored phase (such as FCC in this study) --- can be controlled by adjusting the seed size. While we did not observe the growth of a pure FCC polymorph, the significant increase in intermediate FCC-like particles suggests that understanding the balance between the free energy surface (or, the stability of different polymorphs) and surface-induced local ordering in the fluid phase (which depends on the seed surface morphology and size) may guide us to reliably control the selection of the desired polymorph. This observed strong finite-size effect is particularly relevant to polymorph selection in confined systems, where surface-induced ordering can be modulated by changing the confinement size~\cite{chen2021morphology}.  
\subsection{\label{subsec:6} How does the seed surface morphology and size alter the local structural features of the surrounding fluid?}
Recent studies suggest that the key to polymorph selection lies in the local structural fluctuations of the metastable fluid phase~\cite{tanaka_soft_2012, gispen2023crystal}. To address the question of the signatures of the surface-induced polymorph selection hidden in the metastable fluid phase, we have carefully examined the local structural changes in the metastable fluid phase induced by different types and sizes of pinned seeds. In Fig.~\ref{figure7}A, we present the coarse-grained bond orientational order parameter $\bar{q_6}$ distribution of fluid particles in the vicinity of different (both, crystalline and amorphous) seeds. It is evident from the figure that the interfacial fluid's local structure is highly sensitive to the size and surface morphology of the seed. For seed types with crystalline order (BCC, FCC, and SC), the interfacial fluid becomes increasingly structured as the seed size grows, unlike the amorphous-ordered RS seed, where the interfacial fluid structure remains relatively unaffected by the seed size. Moreover, the enhancement of the interfacial fluid's local structural order is more pronounced for the BCC and FCC seeds compared to the SC and RS seeds (see Figs.~\ref{figure7}A and~\ref{figure7}B).  The observed significant reduction in the nucleation barrier with increasing seed size for the FCC and BCC cases (see Fig.~\ref{figure3}) can be attributed to the surface-induced enhancement of the local structure in the interfacial fluid particles.

We further probed the polymorphic identity of the partially structured fluid particles (particles with $\bar{q_6} > 0.25$) in the interfacial region by calculating the local $\bar{w_6}$ distributions of these particles (see Fig.~\ref{figure7}C ). For the BCC and FCC seeds, the interfacial fluid particles exhibit a polymorphic identity similar to that of the seed, \textit{i.e.}, BCC-like ($\bar{w_6} > 0$) and FCC-like ($\bar{w_6} < 0$), respectively (see Fig.~\ref{figure7}D). Interestingly, in the case of the SC seed, BCC-like local structures, rather than SC-like, are weakly favored. The amorphous-ordered RS seed does not impart any polymorphic bias to the interfacial fluid particles. The enhanced nucleation of the BCC phase in the presence of the BCC seed, and the competition between BCC and FCC phase nucleation with the FCC seed, can be attributed to the surface-induced FCC-like local structural bias within the bulk metastable fluid, where BCC is the most stable phase. Thus, the surface-induced selection of the polymorphic identity in the final crystalline phase can be inferred by examining the local structural order of the interfacial and bulk fluid particles.
\section{Conclusions}\label{conclusions}
In this study, we systematically examined the effects of quenched disorder (or, static impurities) of varying length scales and surface morphologies on the nucleation mechanism and kinetics in the GCM system, which mimics soft colloidal systems and exhibits crystal polymorphism. This makes it ideal for investigating the impact of static impurities on both nucleation kinetics and substrate-induced polymorph selection. We first investigated the sensitivity of the nucleation free energy barrier and the critical cluster size to the fraction of point pinned particles, $f_{\rm p}$.  Both the nucleation free energy barrier and critical cluster size are highly sensitive to $f_{\rm p}$, and increase sharply as  $f_{\rm p}$ is increased.  Additionally, we explored the dependence of the nucleation free energy barrier on the length scale of inhomogeneities (quenched disorder) by randomly freezing regions of different sizes within the system. As the size of the pinned cluster/region increases, the nucleation free energy barrier and the critical cluster size decrease, gradually approaching the corresponding homogeneous limit, with no evidence of heterogeneous nucleation.  

To gain deeper insights into the effects of static impurities on nucleation mechanism and kinetics, we examined the influence of surface morphology of pinned clusters (referred to as ``seeds") of different sizes. The nucleation barrier decreases significantly  with increasing seed size for FCC, BCC, and SC seeds. Among them, BCC seeds facilitated nucleation more effectively than FCC or SC seeds. For both BCC and FCC seeds, we observed a size-dependent crossover from activated to barrierless (heterogeneous) nucleation, where the phase transition becomes entirely downhill. In contrast, the amorphous-ordered RS seed does not facilitate nucleation kinetics. Our findings also showed that the polymorphic identity of particles in the final crystalline phase can be controlled by altering the seed surface morphology. This (surface-induced) selection of the polymorphic identity can be inferred by probing the local structural order of the interfacial and bulk fluid particles in the metastable fluid. We also investigated the system size dependence of the phase transition pathways for the FCC seed case and found that the polymorphic identity of particles in the final crystalline phase is strongly influenced by system size. This strong finite-size effect is particularly relevant to polymorph selection in confined systems, where surface-induced ordering can be tuned by altering the confinement size.

Finally, it is important to note that in this study, impurities interact with fluid particles with the same GCM interaction potential as the fluid particles themselves. Future works could explore the effects of altering impurity-fluid interactions on nucleation kinetics and phase transition pathways~\cite{sosso2022role, qiu2017strength}. Nonetheless, we believe the insights gained from this study will aid in controlling crystallization processes in experiments to achieve the desired polymorph.

\begin{acknowledgments}
R.S.S. acknowledges financial support from DST-SERB (Grant No. SRG/2020/001415 \& CRG/2023/002975). G.R. acknowledges financial support from IISER Tirupati. M. S. acknowledges financial support from DST-SERB (Grant No. SRG/2020/001385). The computations were performed at the IISER Tirupati computing facility and at PARAM Brahma at IISER Pune. 
\end{acknowledgments}

\bibliographystyle{apsrev4-2}
\bibliography{Main}

\pagebreak
\newpage
\widetext
\begin{center}
\textbf{\large Supplementary Materials}
\end{center}
\setcounter{equation}{0}
\setcounter{figure}{0}
\setcounter{table}{0}
\setcounter{page}{1}
\setcounter{section}{0}
\makeatletter
\renewcommand{\theequation}{S\arabic{equation}}
\renewcommand{\thefigure}{S\arabic{figure}}

\begin{figure*} [h]
    \centering
    \includegraphics[scale=0.4]{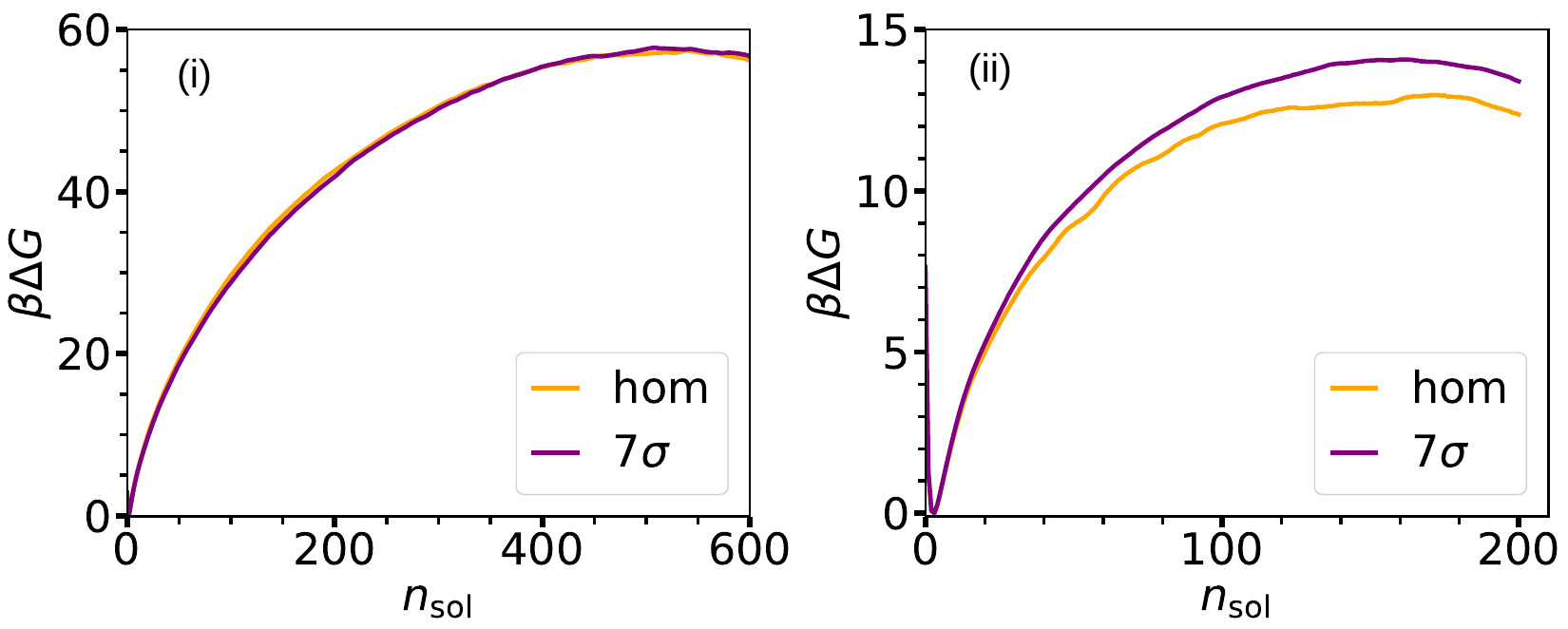}
    \caption{The nucleation free energy profile ($\beta \Delta G$) for the solid-like clusters from the metastable fluid for the homogenoeus case (indicated by ``hom") and in the presence of an amorphous ordered RS seed of radius $7\sigma$ at SP1 (i) and SP2 (ii) is shown for $N = 6000$. It is evident that the rough amorphous surface morphology does not give rise to any nucleation kinetics facilitation over the homogeneous case.}
    \label{figs1}
\end{figure*}

\begin{figure}
    \centering
    \includegraphics[scale=0.5]{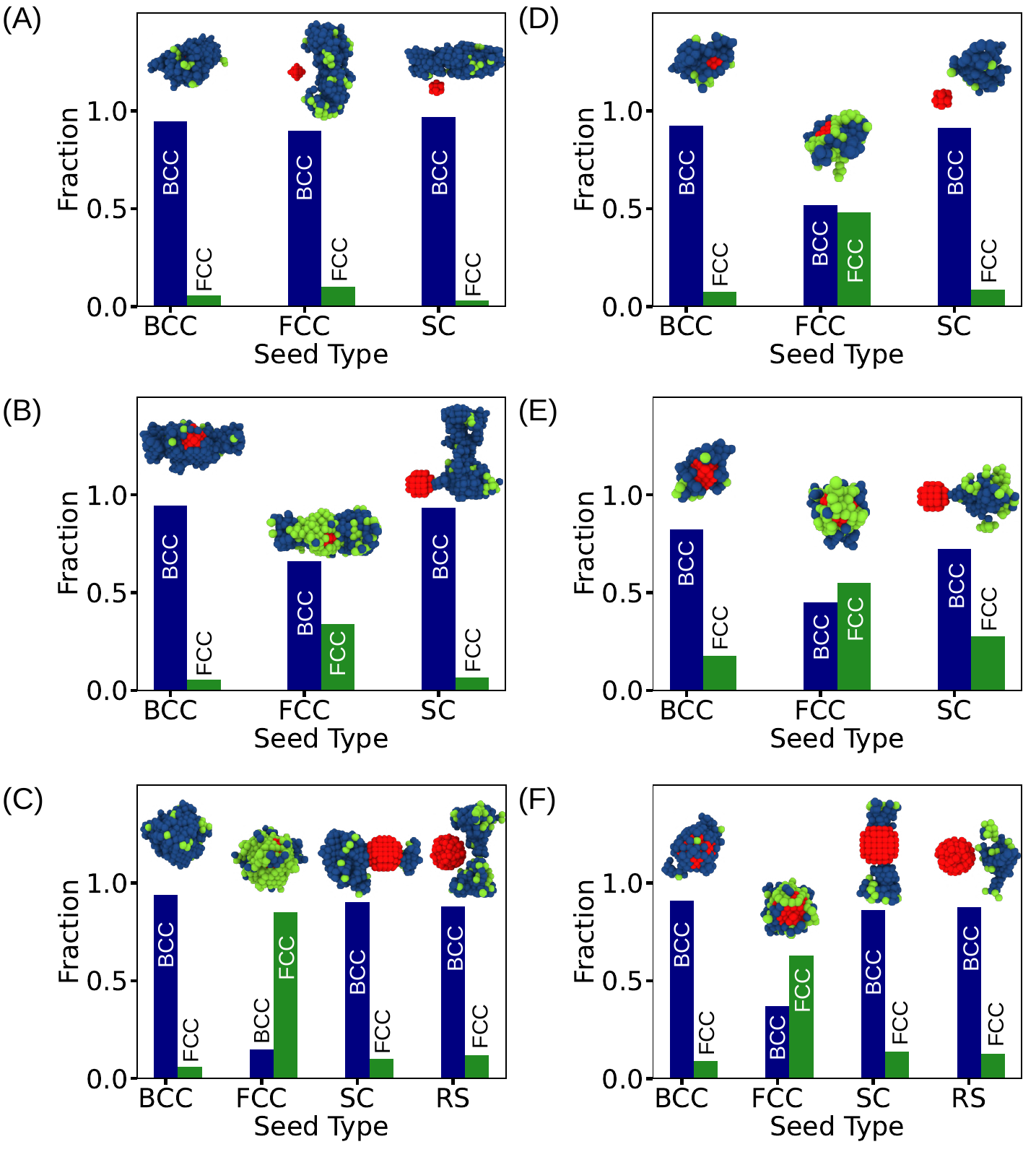}
    \caption{The composition of the critical cluster is shown for seed radius $3\sigma$ at SP1 (A) and SP2 (D), $5\sigma$ at SP1 (B) and SP2 (E), and $7\sigma$ at SP1 (C) and SP2 (F). The seed is represented in red color, the BCC-like particles in blue, and the FCC-like particles in green. Note that the critical cluster size is approximately $600$ at SP1, and $200$ at SP2.}
    \label{figs2}
\end{figure}

\begin{figure}
    \centering
    \includegraphics[scale=0.34]{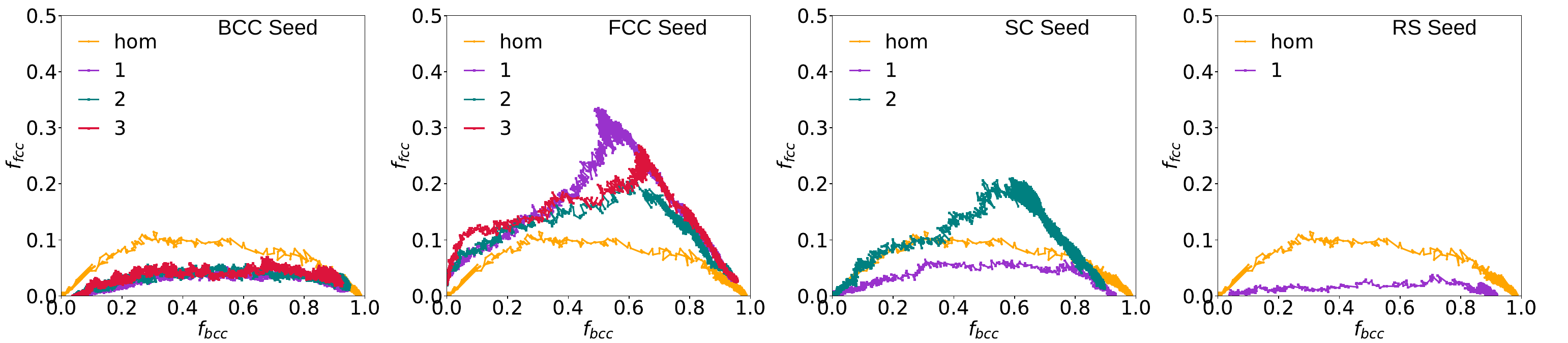}
    \caption{The phase transition pathways of the metastable fluid phase in the two-dimensional ($f_{\rm bcc} - f_{\rm fcc}$) order parameter space at SP2 is shown. The system consists of $N = 10000$ particles along with different types of seeds of radius $7 \sigma$ at the center of the box. The orange line indicates the phase transition path for the homogeneous case.}
    \label{figs3}
\end{figure}

\begin{figure*} [h]
    \centering
    \includegraphics[scale=0.28]{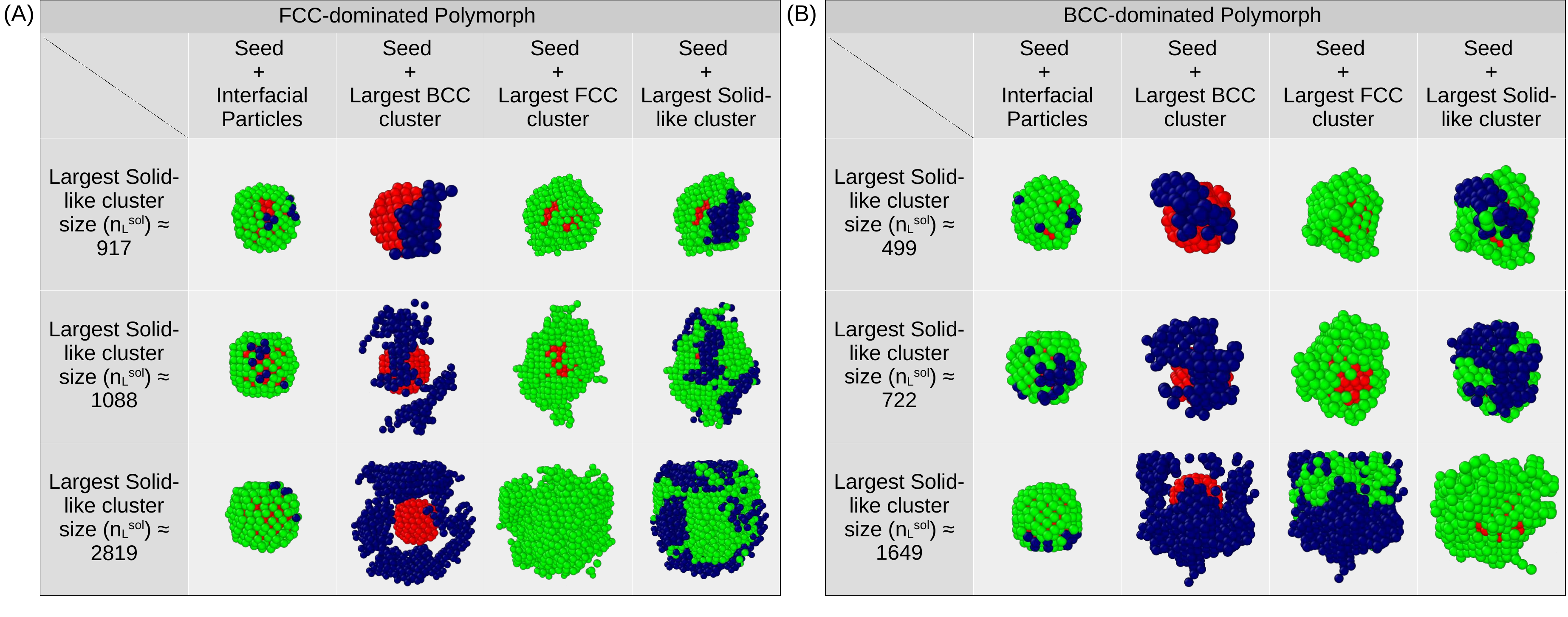}
    \caption{ The representative snapshots during the transition of the metastable fluid into the FCC-dominated phase for $N = 6000$ (A), and the BCC-dominated phase for $N = 10000$ (B) at SP1. Here, we have shown the seed of radius $7\sigma$ along with the polymorphic identity of the interfacial solid-like particles, the largest BCC- and FCC-like clusters, and the largest solid-like cluster during the growth. We note that, for both the cases (A and B), nucleation of the solid-like particles starts on the seed surface. The blue spheres represent the BCC-like and the green spheres represent the FCC-like particles in the system. The seed particles are represented in red.}
    \label{figs4}
\end{figure*}

\end{document}